\documentclass[preprint,12pt]{elsarticle}

\usepackage[margin=1in]{geometry}

\usepackage{graphicx}

\usepackage{amssymb}

\usepackage{url}

\usepackage{amsmath}
\usepackage{upgreek}

\usepackage{amsthm}

\usepackage{tensor}
\usepackage{dsfont}

\usepackage{siunitx}

\usepackage{amssymb}

\usepackage{bbm}

\usepackage{soul}

\usepackage{chemformula}

\usepackage{algorithmicx}
\usepackage{algorithm}
\usepackage[noend]{algpseudocode}
\makeatletter
\renewcommand{\ALG@beginalgorithmic}{\footnotesize}
\makeatother

\usepackage{hyperref}

\usepackage{booktabs}

\usepackage{amsmath}

\makeatletter
\newcommand{\vast}{\bBigg@{4}}
\newcommand{\Vast}{\bBigg@{7}}
\makeatother

\makeatletter
\def\ps@pprintTitle{%
  \let\@oddhead\@empty
  \let\@evenhead\@empty
  \let\@oddfoot\@empty
  \let\@evenfoot\@oddfoot
}
\makeatother

\usepackage{todonotes}

\newcommand*\xbar[1]{%
   \hbox{%
     \vbox{%
       \hrule height 0.5pt 
       \kern0.5ex
       \hbox{%
         \kern-0.1em
         \ensuremath{#1}%
         \kern-0.1em
       }%
     }%
   }%
}

\usepackage{caption}

\usepackage{subcaption}

\setcitestyle{authoryear}
 \biboptions{comma,round}

\date{}

\journal{Journal of Computational Physics}

\begin{document}

\begin{frontmatter}

\title{A model for transport of soluble surfactants in two-phase flows}


\author{Suhas S. Jain\corref{cor1}}
\ead{suhasjain@gatech.edu}

\address{Flow Physics and Computational Science Lab, Georgia Institute of Technology, GA, USA.\\
Center for Turbulence Research, Stanford University, CA, USA.}

\begin{abstract}


In this work, we propose a novel transport model for soluble surfactants in two-phase flows.
In a two-phase flow, the soluble surfactants can adsorb/desorb from/into the bulk of any of the phases to the interface and can modify the interface properties. This results in sharp gradients in the surfactant concentration on the interface and also between the two phases in the bulk when there is selective adsorption/desorption, presenting a serious challenge for the numerical simulations.

To overcome this challenge, we propose a computational model for the transport of soluble surfactants that can model the adsorption and desorption processes accurately.
The model is discretized using a central-difference scheme, which leads to a non-dissipative implementation that is crucial for the simulation of turbulent flows. The model is used with the ACDI diffuse-interface method \citep{jain2022accurate}, but can also be used with other algebraic-based interface-capturing methods. Furthermore, the provable strengths of the proposed model are: (a) the model maintains the positivity property of the surfactant concentration field, a physical realizability requirement for the simulation of surfactants, when the proposed criterion is satisfied, (b) the proposed model maintains discrete confinement of the interfacial and bulk surfactants and prevents artificial numerical diffusion of the surfactant between the interface and the bulk and between the two phases in the bulk.

Finally, we present numerical simulations using the proposed model for both one-dimensional and multi-dimensional cases and assess: the accuracy and robustness of the model, the validity of the positivity property of the scalar concentration field, and the confinement of the surfactant at the interface. We also study the effect of surfactants on an oscillating droplet and on a complex droplet/bubble-laden turbulent flow.

\end{abstract}


\end{frontmatter}



\section{Introduction} 


Surfactants modulate surface tension properties and generate Marangoni forces, and as such, they find diverse applications in multiphase flows across various industries. In the oil and gas sector, they are used for stabilizing emulsions for efficient pipeline transport and to aid in breaking them during oil separation. Chemical-enhanced oil recovery utilizes surfactants to alter interfacial tension, improving oil displacement from reservoirs \citep{massarweh2020use}. In pharmaceuticals and agrochemicals, surfactants create stable microemulsions to enhance solubility \citep{castro2014advances}. Environmental applications involve surfactants in bioremediation, aiding contaminant removal \citep{churchill1995surfactant}. Surfactants play crucial roles in personal care product formulation \citep{rhein2006surfactants}, food industry emulsion stability \citep{kralova2009surfactants}, and drug delivery systems \citep{lawrence1994surfactant}. They also contribute to enhanced gas-liquid mass transfer in bioprocess engineering and microfluidic manipulation for lab-on-a-chip applications \citep{baret2012surfactants}, showcasing their versatility in optimizing multiphase flow processes.



The effect of surfactants on two-phase flows has been studied extensively using various approaches, including experimental, theoretical, semi-analytical, and computational techniques, such as boundary integral-based methods as well as sharp-interface- and diffuse-interface-based continuum approaches. 

Some of the recent advancements in modeling insoluble surfactants include the volume-of-fluid (VOF)-based formulation presented by \citet{james2004surfactant}, finite-element-based methods by \citet{venkatesan2019simulation} and \citet{frachon2023cut}, segment projection method by \cite{khatri2011numerical}, level-set method by \cite{xu2012level}, front-tracking method with adaptive mesh refinement by \cite{de20153d}, immersed-boundary method by \cite{lai2008immersed}, hybrid methods by \cite{cui2011computational}, Cahn-Hilliard-based diffuse interface methods by \citet{abels2019existence,di2022well,teigen2009diffuse,teigen2011diffuse,garcke2014diffuse}, and \citet{ray2021discontinuous}, and second-order phase-field-based method by \citet{jain2024model}.

Similarly, soluble surfactants have been modeled extensively, e.g., using finite-element-based formulation with a coupled arbitrary Lagrangian-Eulerian and Lagrangian approach for modeling interfaces by \citet{ganesan2012arbitrary}, a parametric finite-element approximation for interface and surface finite-element approximation for surfactants by \citet{barrett2015stable}, a segment projection method by \citet{khatri2014embedded}, a level-set method by \citet{xu2018level}, a front-tracking method by \citet{muradoglu2008front}, an immersed-boundary method by \citet{chen2014conservative}, Cahn-Hilliard-based diffuse-interface methods by \citet{teigen2009diffuse,teigen2011diffuse,soligo2019coalescence,bau2025transfer}, lattice-Boltzmann methods by \citet{liu2010phase} and \citet{kothari2023free}, and an algebraic volume-of-fluid method by \citet{antritter2024two}. 
Other advancements in modeling soluble surfactants include a multiscale modeling approach presented by \citet{booty2010hybrid}, porous media flows by \citet{zhang2021pore}, and linear stability analysis of effect of soluble surfactants on two-phase flows by \citet{herrada2022effect}. However, to the best of our knowledge, there is no model for transport of soluble surfactants for second-order phase-field methods.

We recently developed a model for transport of interface-confined scalars and insoluble surfactants \citep{jain2024model} and used it with the accurate conservative diffuse-interface (ACDI) method, a second-order phase-field method. However, this model is not limited to a diffuse-interface approach; it can be used with any other interface-capturing method. 
The primary objective of the present work is to extend the previous work in \citet{jain2024model} for modeling transport of soluble surfactants by coupling it with a transport model for surfactants in the bulk, and by incorporating adsorption and desorption physics of the surfactants at the interface. The proposed model is also capable of accurately modeling selective adsorption/desorption from/into one of the phases without resulting in leakage between the phases or between the bulk and the interface.
The proposed model does not require division by $\phi$ (volume fraction), and therefore doesn't require any special treatment when $\phi$ goes to $0$.
We prove and show that the total concentration of the surfactant remains positive and conserved, which is a physical-realizability condition, using second-order central-difference schemes. The second-order central-difference schemes are chosen due to their low-dissipative nature that are known to be suitable for simulations of complex turbulent flows \citep{moin2016suitability}.

We use a second-order phase-field method, particularly the ACDI method by \cite{jain2022accurate}, to model the interface in a two-phase flow. 
The proposed model can also be used with a conservative phase-field/diffuse-interface method \citep{chiu2011conservative}, 
a conservative level-set method \citep{olsson2005conservative}, an accurate conservative level-set method \citep{Desjardins2008}, including the compressible variant of the diffuse-interface methods with four-equation \citep{jain2023assessment}, five-equation \citep{jain2020conservative}, or the six-and seven-equation models \citep{hatashita2025interface} and any other method that results in a hyperbolic tangent interface shape in equilibrium, and when the volume fraction $\phi$ is bounded between $0$ and $1$. For coupling with other models, like a Cahn-Hilliard model where the volume fraction takes values between $-1$ and $1$, the proposed model can be affine transformed with respect to the order parameter, such that the change in the range from $[0,1]$ to the range of values of $\phi$ that the interface-capturing model admits is accounted for.  

We present numerical simulations of transport of soluble surfactants with different solubilities in the two phases to illustrate the accuracy, consistency, and robustness of the proposed method in both simple settings and complex turbulent flows.

\subsection{Sharp vs diffuse representations}

\begin{figure}
    \centering
    \includegraphics[width=0.9\linewidth]{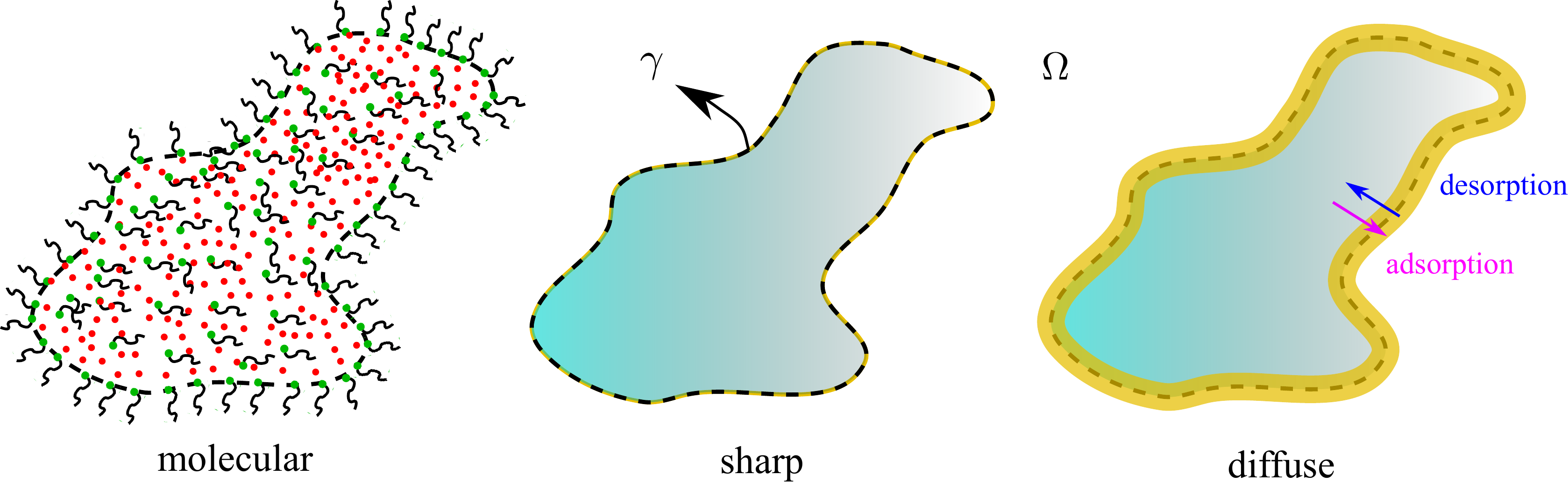}
    \caption{A schematic showing a soluble surfactant in molecular and continuum (sharp and diffuse) representations. Here, $\gamma$ represents the two-dimensional interface embedded in a three-dimensional domain $\Omega$, where the dashed line represents the interface. In the molecular picture (not drawn to scale), the dispersed phase molecules are shown along with the surfactant molecules that are adsorbed on the interface and in the bulk of the dispersed phase. In the continuum representations, the colored solid line represents surfactant concentration on the interface, and the gradient filled color represents surfactant concentration in the bulk.}
    \label{fig:schematic-2D}
\end{figure}

Consider the schematic of an interface $\gamma$ in a domain $\Omega$ shown in Figure \ref{fig:schematic-2D} in molecular and continuum (sharp and diffuse) representations. The molecular picture shows the surfactant molecules adsorbed onto the interface and also in the bulk of the dispersed phase. 
There is also an exchange of the surfactant molecules between the bulk and the interface through adsorption and desorption processes. 
Moreover, there can be selective adsorption/desorption to/from one of the phases, as illustrated in Figure \ref{fig:schematic-2D} where the surfactant is only dissolved in the dispersed phase and not in the carrier phase.

Both the interface and the layer of the surfactant molecules on the interface are typically $O(nm)$ thick. Hence they can be mostly modeled as sharp quantities in macroscopic continuum representation. 
We can write the evolution equations for soluble surfactants (in a sharp representation) as 
\begin{equation}
\frac{\partial \hat{c}_i}{\partial t} + \vec{u}_s\cdot \vec{\nabla}_s \hat{c}_i = \vec{\nabla}_s \cdot \left(D_i \vec{\nabla}_s \hat{c}_i \right) - \hat{c}_i\vec{\nabla}_s\cdot \vec{u}_s - \hat{c}_i \kappa \vec{u}\cdot\vec{n} + \sum_l \hat{j}_l, 
\label{equ:sharp_interface_model_1}
\end{equation}
\begin{equation}
\frac{\partial \tilde{c}_{b,l}}{\partial t} + \vec{\nabla}\cdot(\vec{u} \tilde{c}_{b,l}) = \vec{\nabla} \cdot (D_{b,l} \vec{\nabla} \tilde{c}_{b,l}) - \hat{j}_l ,    
\label{equ:sharp_bulk_model_1}
\end{equation}
where 
$\hat{c}_i$ represents the surfactant concentration on the interface per unit area and $\tilde{c}_{b,l}$ represents the surfactant concentration in the bulk per unit volume of the phase $l$;
$\vec{\nabla}_s=(I-\vec{n}\vec{n})\vec{\nabla}$ is the surface gradient; $\vec{n}$ is the interface normal; $\kappa$ is the curvature; $\vec{u}_s=(I-\vec{n}\vec{n})\vec{u}$ is the surface velocity; and the source/sink term that arises due to species adsorption into and desorption out of the interface \citep{martinez2020langmuir} is 
\begin{equation}
\hat{j}_l=r_{a,l} \tilde{c}_{b,l} (\hat{c}_{i,\infty} - \hat{c}_i) - r_{d,l} \hat{c}_i.    
\end{equation}
Surfactant concentration will saturate when $\hat{c}_i$ reaches $\hat{c}_{i,\infty}$, the critical concentration, and hence the form of $\hat{j}_l$. 
Here, $r_{a,l}$ is the rate of adsorption of surfactant from the bulk of phase $l$ into the surface and $r_{d,l}$ is the rate of desorption of surfactant from the surface to the bulk of phase $l$; $D_i$ and $D_{b,l}$ are the interfacial and bulk diffusivity of the surfactant, respectively.

However, this sharp nature of the surfactant layer poses a challenge in the numerical simulations on an Eulerian grid because of the sharp jump in the concentration values at the interface. In addition to this, the selective adsorption and desorption into one of the phases can also result in a sharp jump in surfactant concentration values in the bulk on the two sides of the interface. If not treated carefully, these could result in negative concentration values and artificial numerical leakage of the surfactant from the interface to the bulk and across the interface from one phase to the other.

To overcome these challenges, we numerically diffuse the surfactant concentrations in the interface normal direction, as shown in Figure \ref{fig:schematic-2D}, in such a way that the gradients in the concentration can now be resolved on an Eulerian grid. 
With this, we could construct corresponding diffuse quantities for $\hat{c}_i$ and $\tilde{c}_{b,l}$ as $c_i=\delta_s \hat{c}_i$ (interfacial concentration per unit volume) and $c_{b,l} = \phi \tilde{c}_{b,l}$ (bulk concentration per unit volume), where $\delta_s$ is a surface delta function, defined as $\int_{\gamma} \hat{c}_i\ d\gamma = \int_{\Omega} \hat{c}_i \delta_s\ d\Omega$. 

\section{Phase-field model}


 
In this work, we use the ACDI method by \cite{jain2022accurate}, which is an Allen-Cahn-based second-order phase-field model given by
\begin{equation}
\frac{\partial \phi}{\partial t} + \vec{\nabla}\cdot(\vec{u}\phi) = \vec{\nabla}\cdot\left\{\Gamma\left\{\epsilon\vec{\nabla}\phi - \frac{1}{4} \left[1 - \tanh^2{\left(\frac{\psi}{2\epsilon}\right)}\right]\frac{\vec{\nabla} \psi}{|\vec{\nabla} \psi|}\right\}\right\},
\label{eq:ACDI}
\end{equation} 
where $\phi$ is the phase-field variable that represents the volume fraction of phase $l$ ($\phi = \phi_l$), $\vec{u}$ is the velocity, $\Gamma$ represents the velocity-scale parameter, $\epsilon$ is the interface-thickness-scale parameter, and $\psi$ is an auxiliary signed-distance-like variable given by 
\begin{equation}
    \psi = \epsilon \ln\left(\frac{\phi + \varepsilon}{1 - \phi + \varepsilon}\right),
    \label{eq:psi}
\end{equation} 
where $\varepsilon=10^{-100}$ is a small number. The parameters are chosen to be $\Gamma \ge |\vec{u}|_{max}\ \mathrm{and}\ \epsilon > 0.5 \Delta x$, and $\Delta t$ satisfies the explicit Courant-Friedrichs-Lewy criterion, to maintain the boundedness of $\phi$ \citep{jain2022accurate}.
The ACDI model is more accurate than other phase-field models because it maintains a sharper interface (with only one-to-two grid points across the interface) while being robust and conservative, without the need for any geometric treatment. It has recently been extended to an unstructured framework for simulations in complex geometries \citep{hwang2024robust} and other multiphysics applications, such as modeling solidification \citep{brown2023phase} and boiling \citep{scapin2022boil} in two-phase flows. Hence, the ACDI method is chosen as the interface-capturing method in this work. 

To model surface tension forces, we use the improved continuum surface force (CSF) approach in \citet{jain2022accurate} where $\psi$ from Eq. \eqref{eq:psi} is used to compute the curvature and normals in the surface tension force. This was shown to result in two-orders of magnitude lower spurious currents and improved accuracy compared to the classical CSF model.

 

\section{Proposed model for the transport of soluble surfactants in two-phase flows \label{sec:proposed-model}}

We propose to model soluble surfactants using three transport equations ``a three-equation model", one equation for the transport of the interfacial surfactant concentration, $c_i$, and two more for the bulk surfactant concentration, $c_{b,l}$, in each of the two phases $l$, along with source/sink terms for the exchange between the interfacial and bulk equations. Two separate equations are chosen for transport in the bulk, one for each of the phases, to accurately model the selective adsorption/desorption of the surfactant to each of the phases.
Both $c_i$ and $c_{b,l}$ are volumetric quantities, which represent concentration as the amount of species per unit volume. Accordingly, $\tilde{c}_{b,l}=c_{b,l}/\phi$ is the local bulk concentration of surfactant defined as the amount of species per unit volume of the phase $l$, and $\hat{c}_i=c_i/\delta_s$ is the local interfacial concentration of surfactant defined as the amount of species per unit interfacial area, where $\delta_s=|\vec{\nabla}\phi|$ is the surface delta function.

The proposed model for the transport of soluble surfactants in two-phase flows is 
\begin{equation}
\frac{\partial c_i}{\partial t} + \vec{\nabla}\cdot(\vec{u} c_i ) = \vec{\nabla} \cdot \left[D_i \left\{\vec{\nabla}c_i - \frac{2(0.5 - \phi) \vec{n} c_i}{\epsilon} \right\}\right] + \sum_l \hat{j}_l \delta_s ,
\label{equ:interface_surfactant_model}
\end{equation}
\begin{equation}
\frac{\partial c_{b,l}}{\partial t} + \vec{\nabla}\cdot(\vec{u} c_{b,l}) = \vec{\nabla} \cdot \left[D_{b,l} \left\{\vec{\nabla}c_{b,l} - \frac{(1 - \phi) \vec{n} c_{b,l}}{\epsilon} \right\}\right] - \hat{j}_l \delta_s ,
\label{equ:bulk_surfactant_model}
\end{equation}
where $\vec{n}=\vec{\nabla}\phi/|\vec{\nabla}\phi|=\vec{\nabla}\psi/|\vec{\nabla}\psi|$ is the interface normal vector. 

\begin{figure}
    \centering
    \includegraphics[width=0.65\textwidth]{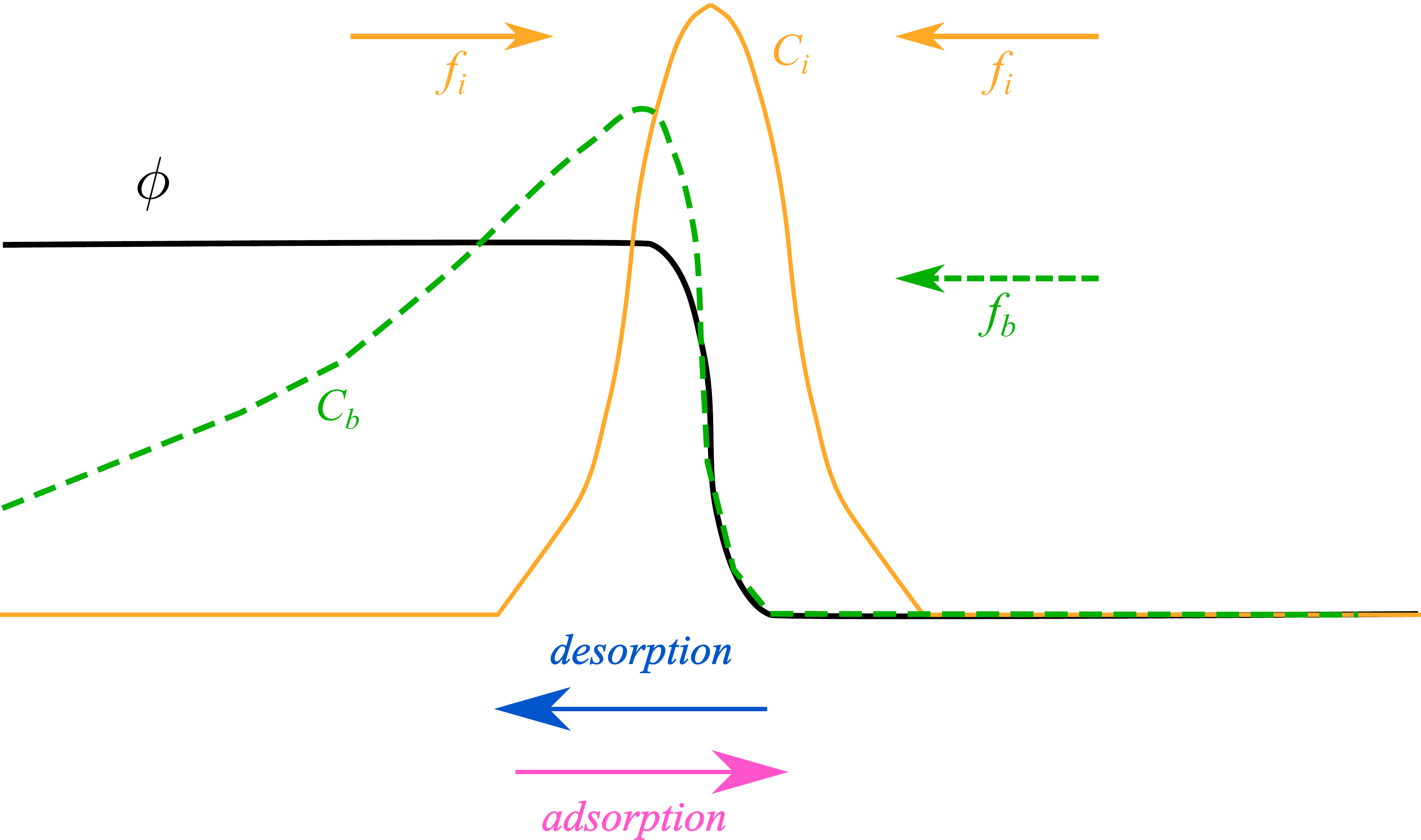}
    \caption{Schematic representing the interfacial surfactant concentration, $c_i$, the bulk surfactant concentration, $c_b$, and the exchange between these two due to adsorption and desorption of the surfactant. The effect of the sharpening flux terms $f_i$ and $f_b$ in the model are shown. Here, the surfactant is only dissolvable in the bulk of the phase represented by $\phi=1$.}
    \label{fig:schematic}
\end{figure}

The second term on the right-hand side of Eqs. \eqref{equ:interface_surfactant_model} and \eqref{equ:bulk_surfactant_model} are artificial sharpening terms $f_i=D_i 2(0.5 - \phi)\vec{n}c_i/\epsilon$ and $f_b=D_{b,l} (1 - \phi) \vec{n}c_{b,l}/\epsilon$. The effect of $f_i$ is to prevent the artificial diffusion of the interfacial surfactant on both sides of the interface so that it is confined to the interface region, and $f_i$ switches sign for values below and above $\phi=0.5$ as shown in Figure \ref{fig:schematic}.
The effect of $f_b$ is to prevent the diffusion of the bulk surfactant into the other phase, and as such $f_b$ is only active when $\phi$ is not equal to $1$ as shown in Figure \ref{fig:schematic}. Additional consistent terms and alternate model forms can be derived (Appendix A) for the proposed model; however, these terms are in non-conservative form and could also make the model non-robust due to the need for division by $\phi$ and $\delta_s$. Therefore, the proposed model in Eqs. \eqref{equ:interface_surfactant_model} and \eqref{equ:bulk_surfactant_model} is preferred over the ones in Appendix A.



\subsection{Two-way coupling}

To account for two-way coupling with hydrodynamics, this work uses a linearized version of the Langmuir equation of state (EOS), which relates the surfactant concentration to the surface tension coefficient. The Langmuir EOS \citep{tricot1997surfactants} can be written as
\begin{equation}
    \sigma(\hat{c}_i) = \sigma_0 \left[1 + \frac{R T \hat{c}_{i,\infty}}{\sigma_0} \ln{\left(1 - \frac{\hat{c}_i}{\hat{c}_{i,\infty}}\right)} \right], 
\end{equation}
where $R$ is the ideal gas constant, $T$ is the absolute temperature, $\hat{c}_{i,\infty}$ is the maximum interfacial surfactant concentration, and $\sigma_0$ is the surface tension for the clean interface. In the low surfactant concentration limit, this can be reduced to a linear model (also called Henry's EOS) as
\begin{equation}
    \sigma(\hat{c}_i) = \sigma_0 \left( 1 - Ma \frac{\hat{c}_i}{\hat{c}_{i,\infty}} \right),
\end{equation}
where $Ma=R T \hat{c}_{i,\infty}/\sigma_0$ is the Marangoni elasticity number, which is a measure of sensitivity of the surface tension to the surfactant concentration.

With a varying surface tension coefficient, the surface tension force can be modeled as \citep{landau2013fluid}
\begin{equation}
    \vec{F}_{\sigma} = \sigma \kappa \vec{\nabla}\phi - (\vec{\nabla} \sigma) |\vec{\nabla}\phi|, 
\end{equation}
where the first term is the capillary force and the second term is the Marangoni force.

\section{Positivity \label{sec:positivity_proof}}

In the absence of exchange of the surfactant between the bulk and the interface, Eq. \eqref{equ:interface_surfactant_model} reduces to a transport model for an interface-confined scalar \citep{jain2024model}, and Eq. \eqref{equ:bulk_surfactant_model} reduces to a transport model for an immiscible scalar \citep{jain2023scalar}.
In these settings, $c_i$ and $c_{b,l}$ are shown to remain positive if 
\begin{equation}
    \Delta x \le \left(\frac{2 D}{|u|_{\mathrm{max}} + \frac{D}{\epsilon}}\right)
    \label{eq:crossover}
\end{equation}
and 
\begin{equation}
      \Delta t \le \frac{\Delta x ^2}{2N_dD}
    \label{eq:boundtime}
\end{equation}
are satisfied, where $D$ represents $D_i$ for $c_i$ and $D_{b,l}$ for $c_{b,l}$, $\Delta x$ is the grid-cell size, $\Delta t$ is the time-step size, $|u|_{\mathrm{max}}$ is the maximum fluid velocity in the domain, and $N_d$ is the number of dimensions. Note that this also requires $\phi^k_i$ to be bounded between $0$ and $1$, $\forall k\in\mathds{Z}^+$ and $\forall i$, which is guaranteed to be satisfied with the ACDI method \citep{jain2022accurate}. If $\epsilon=\Delta x$, then the constraint in Eq. \eqref{eq:crossover} reduces to
\begin{equation}
\Delta x\le \frac{D}{|u|_{\mathrm{max}}}\ \text{or}\ Pe_c\le1,
\label{eq:positive}
\end{equation}
where $Pe_c=\Delta x |u|_{\mathrm{max}}/D$ is the cell P\'eclet number. Similarly, for $\epsilon=0.75\Delta x$ the constraint is $Pe_c\le0.67$, and for $\epsilon=0.6\Delta x$ the constraint is $Pe_c\le0.33$.

Note that, when we sum up Eqs. \eqref{equ:interface_surfactant_model} and \eqref{equ:bulk_surfactant_model}, we obtain a transport equation for total surfactant concentration $c_i + \sum_l c_{b,l}$, where the source/sink terms in these equations cancel out exactly. Therefore, it is easy to see that the conditions in Eqs. \eqref{eq:crossover} or \eqref{eq:positive} and \eqref{eq:boundtime} are sufficient to maintain the positivity of the total surfactant concentration field.

\section{Numerical methods}

The proposed models are implemented in the GPU-accelerated \href{https://github.com/Flow-Physics-Computational-Science-Lab}{ExaFlow CFD solver}, developed by the Flow Physics and Computational Science Lab at Georgia Tech. The solver can handle both incompressible and compressible flows. For incompressible flows, a finite-volume discretization strategy on a staggered grid has been employed wherein the phase field, the pressure field, and the surfactant concentration fields are stored at the cell centers; and the components of the velocity field vector are stored at the cell faces where all the fluxes are evaluated. This choice of discretization is adopted, for incompressible flows, to avoid the spurious checkerboarding of the pressure field \citep{patankar1980numerical}. The pressure-Poisson equation is solved using the HYPRE package \citep{falgout2002hypre}.  

In this work, we use a second-order central scheme for spatial discretization and a fourth-order Runge-Kutta scheme for time stepping for the proposed model in Section \ref{sec:proposed-model}. A skew-symmetric-like flux-splitting approach \citep{jain2022kinetic} is adopted for the discretization of the ACDI method in Eq. \eqref{eq:ACDI}.


\section{Simulation results}

In this section, the proposed model for soluble surfactant is used to simulate surfactant transport in two-phase flows and the results are presented for wide range of scenarios. Throughout the section, all the parameters are presented in arbitrary simulation units.
The presented simulations can be subdivided into (a) one-dimensional cases with verification of positivity and robustness of the method for simulating adsorption/desorption and selective adsorption/desorption in Section \ref{sec:1D}, and (b) multi-dimensional cases involving adsorption on a stationary droplet, and the effect of surfactants on an oscillating droplet and a complex droplet/bubble-laden turbulent simulation in Section \ref{sec:multiD}.

\subsection{One-dimensional simulations\label{sec:1D}}

In this section, one-dimensional simulations are presented, which act as verification of the proposed model. In all the simulations, a unit domain length of $L=1$ is used with a grid size of $\Delta x=0.01$. A drop of radius $R=0.25$ is initially placed in the domain centered at $x_c=0.5$. The initial condition for the drop is given by $\phi_o = 0.5\left[1 - \tanh{\left\{\left(|x - 0.5| - 0.25\right)/(2\epsilon)\right\}}\right]$ at time $t=0$. 

\subsubsection{Adsorption and desorption \label{sec:adsorp-desorp}}

\begin{figure}[H]
    \centering
    \includegraphics[width=\linewidth]{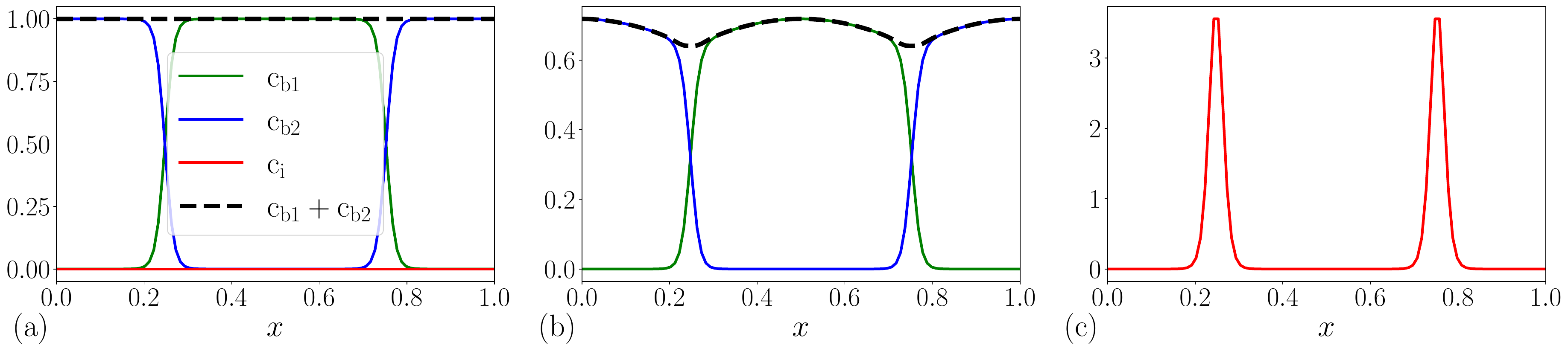}
    \caption{One-dimensional simulation of adsorption of surfactant onto a stationary droplet interface, showing (a) initial bulk and interfacial concentrations at $t=0$, (b) final bulk concentrations at $t=1$, and (c) final interfacial concentration at $t=1$.}
    \label{fig:1D-adsorp}
\end{figure}

To verify the effectiveness of the adsorption and desorption of the surfactant between the bulk and the interface, we simulate two cases of a stationary droplet with soluble surfactants. In the first case, to test the adsorption model, the surfactant is initialized uniformly in the bulk in both phases and there is no initial surfactant on the interface. Here, the parameters are set as $r_{a,l}=1, r_{d,l}=0, D_{b,l}=1,$ and $D_i=1$, and the initial concentration values are set as $c_{b,l} = \phi_l$ and $c_i=0$ at time $t=0$ as shown in Figure \ref{fig:1D-adsorp} (a). At time $t=1$,  the surfactant concentrations in the bulk and on the interface are shown in Figures \ref{fig:1D-adsorp} (b,c). As excepted, the surfactant has adsorbed onto the interface and the bulk concentration has reduced with the concentration values being lowest close to the interface. The surfactant concentration in the bulk on the two sides of the interface are symmetric because the values of $r_{a,l}$ and $D_{b,l}$ are equal for both phases.


\begin{figure}[H]
    \centering
    \includegraphics[width=\linewidth]{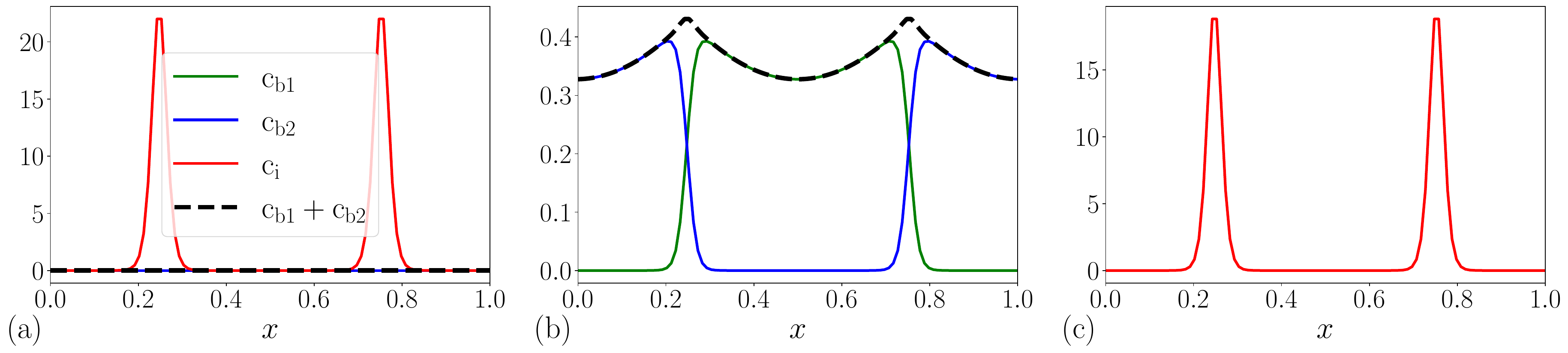}
    \caption{One-dimensional simulation of desorption of surfactant from a stationary droplet interface into the bulk, showing (a) initial bulk and interfacial concentrations, (b) final bulk concentrations at $t=1$, and (c) final interfacial concentration at $t=1$.}
    \label{fig:1D-desorp}
\end{figure}

For the second case, to test the desorption model, the surfactant is initialized only on the interface. The parameters are set as $r_{a,l}=0, r_{d,l}=1, D_{b,l}=1,$ and $D_i=1$, and the initial concentration values are set as $c_{b,l} = 0$ and $c_i=|\vec{\nabla}\phi|$ at time $t=0$ as shown in Figure \ref{fig:1D-desorp} (a). At time $t=1$,  the surfactant concentrations in the bulk and on the interface are shown in Figure \ref{fig:1D-desorp} (b,c). As excepted, a portion of the surfactant has dissolved into the bulk and the interfacial concentration has reduced. The surfactant concentration in the bulk on the two sides of the interface are also symmetric in this case because the values of $r_{d,l}$ and $D_{b,l}$ are equal for both phases.

\subsubsection{Selective adsorption and desorption}
Surfactants can selectively adsorb/desorb to/from interface from one of the phases in the bulk. This can be numerically challenging to simulate because of the jumps in the bulk concentration values at the interface, which can result in artificial numerical leakage of the surfactants from one phase to the other, if not handled carefully. To assess the ability of the proposed method to handle this, we simulate two cases with selective adsorption and two cases with selective desorption of soluble surfactants in a stationary droplet. The diffusivities are chosen as $D_{b,l}=1$ and $D_i=1$ in this section. 

\begin{figure}[H]
    \centering
    \includegraphics[width=\linewidth]{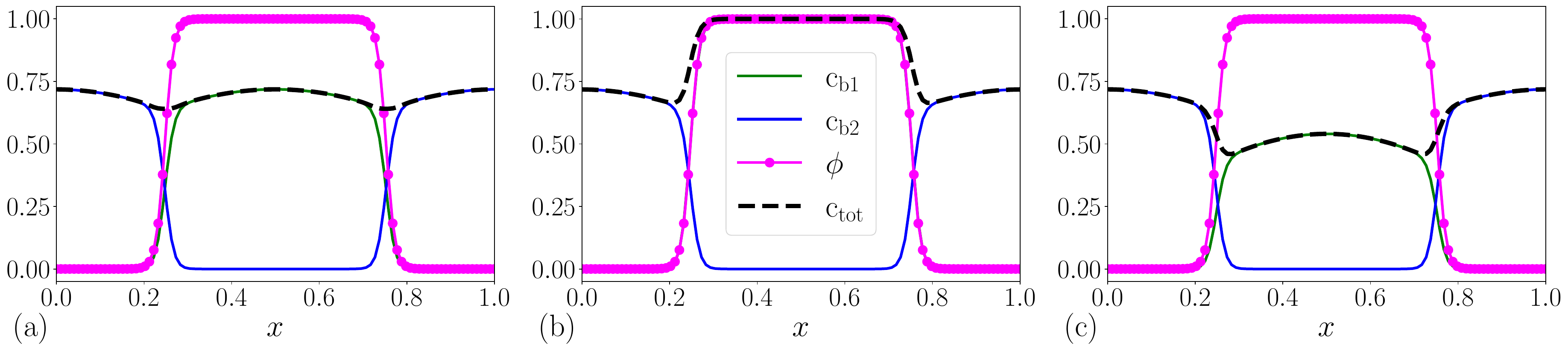}
    \caption{One-dimensional simulation of selective adsorption of surfactant from the bulk to a stationary droplet interface, showing the final time bulk concentrations at $t=1$, for the (a) symmetric case with $r_{a,l} = 1$, (b) non-symmetric case with $r_{a,1} = 0$, $r_{a,2} = 1$ (case 1), and (c) non-symmetric case with $r_{a,1} = 2$, $r_{a,2} = 1$ (case 2).}
    \label{fig:1D-adsorp-select}
\end{figure}

For the pure adsorption cases ($r_{d,l} = 0$), the surfactant is initialized uniformly in the bulk in both phases and there is no initial surfactant on the interface. Here, the adsorption rates are set as $r_{a,1} = 0$, $r_{a,2} = 1$ (case 1) and $r_{a,1} = 2$, $r_{a,2} = 1$ (case 2) for the two cases, respectively, and the initial concentration values are set as $c_{b,l} = \phi_l$ and $c_i=0$ at time $t=0$ as shown in Figure \ref{fig:1D-adsorp} (a). At time $t=1$,  the surfactant concentrations in the bulk are shown in Figures \ref{fig:1D-adsorp-select} (b,c) for the two cases. As excepted, in case 1, the bulk concentration has reduced only for phase 2 whereas the bulk concentration is unchanged for the phase 1 because $r_{a,1} = 0$. In case 2, the bulk concentration for both phases have reduced but the concentration in phase 1 is lower than in phase 2 because of the higher value of adsorption rate in phase 1. For the sake of comparison, the simulation results for $r_{a,l} = 1$ are shown in Figure \ref{fig:1D-adsorp-select} (a) where the bulk concentration is symmetric on both sides of the interface. There was no artificial leakage of the bulk surfactant observed in either of the cases.


\begin{figure}[H]
    \centering
    \includegraphics[width=\linewidth]{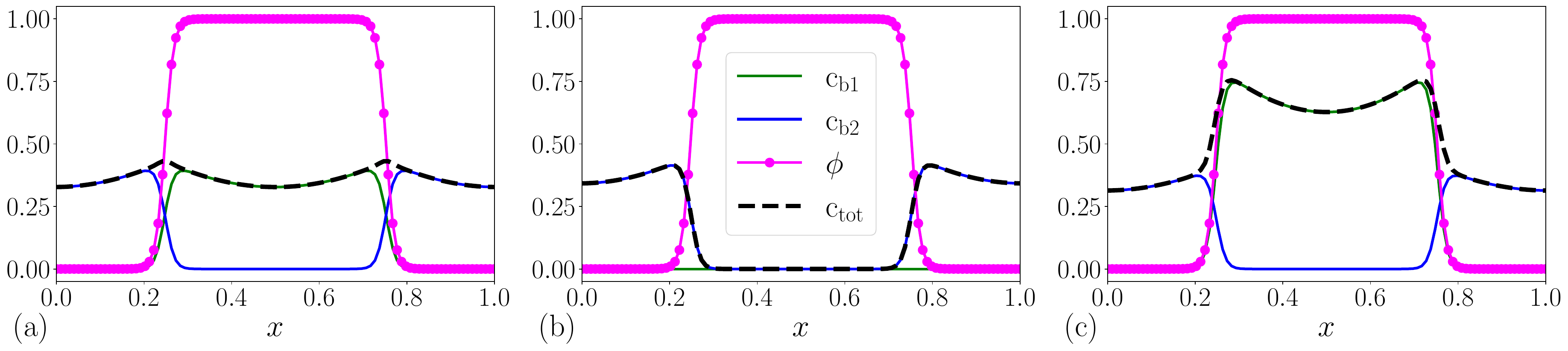}
    \caption{One-dimensional simulation of selective desorption of surfactant from the interface to the bulk in a stationary droplet, showing the final time bulk concentrations at $t=1$, for the (a) symmetric case with $r_{d,l} = 1$, (b) non-symmetric case with $r_{d,1} = 0$, $r_{d,2} = 1$ (case 3), and (c) non-symmetric case with $r_{d,1} = 2$, $r_{d,2} = 1$ (case 4).}
    \label{fig:1D-desorp-select}
\end{figure}

For the pure desorption cases ($r_{a,l} = 0$), the surfactant is initialized on the interface and there is no initial surfactant in the bulk. Here, the desorption rates are set as $r_{d,1} = 0$, $r_{d,2} = 1$ (case 3) and $r_{d,1} = 2$, $r_{d,2} = 1$ (case 4) for the two cases, respectively, and the initial concentration values are set as $c_{b,l} = 0$ and $c_i=c_i=|\vec{\nabla}\phi|$ at time $t=0$ as shown in Figure \ref{fig:1D-desorp} (a). At time $t=1$,  the surfactant concentrations in the bulk are shown in Figures \ref{fig:1D-desorp-select} (b,c) for the two cases. As excepted, in case 3, the bulk concentration is non-zero for phase 2 whereas the bulk concentration is zero for the phase 1 because $r_{a,1} = 0$. In case 4, the bulk concentration for both phases are non-zero but the concentration in phase 1 is higher than in phase 2 because of the higher value of desorption rate in phase 1. For the sake of comparison, the simulation results for $r_{d,l} = 1$ are shown in Figure \ref{fig:1D-desorp-select} (a) where the bulk concentration is symmetric on both sides of the interface. There was again no artificial leakage of the bulk surfactant observed in either of the cases, verifying the robustness of the model.



\subsubsection{Positivity verification}

\begin{figure}
    \centering
    \includegraphics[width=\linewidth]{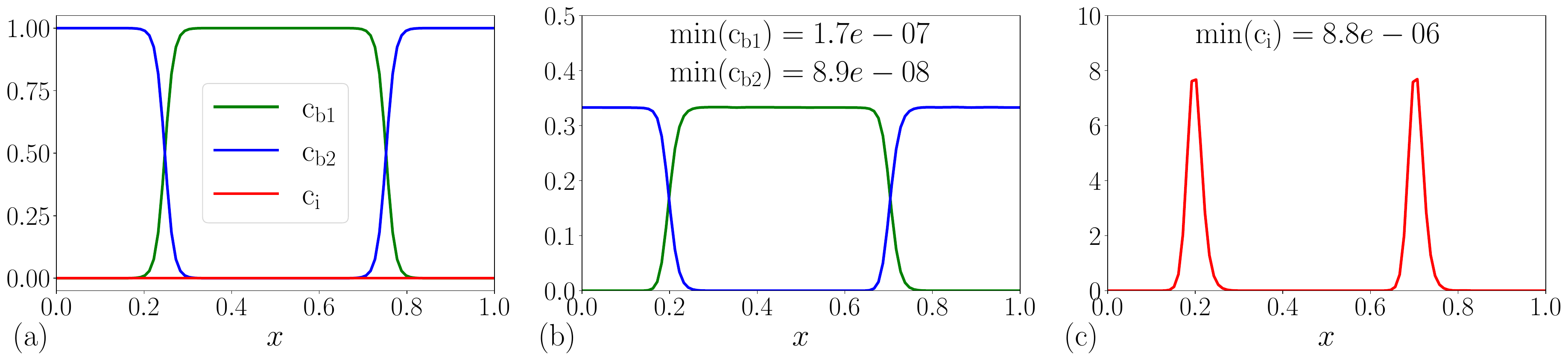}
    \caption{One-dimensional simulation of adsorption of surfactant onto a moving droplet interface for $Pe_c=1$, showing (a) initial bulk and interfacial concentrations at $t=0$, (b) final bulk concentrations, and (c) final interfacial concentration.}
    \label{fig:positive1}
\end{figure}

\begin{figure}
    \centering
    \includegraphics[width=\linewidth]{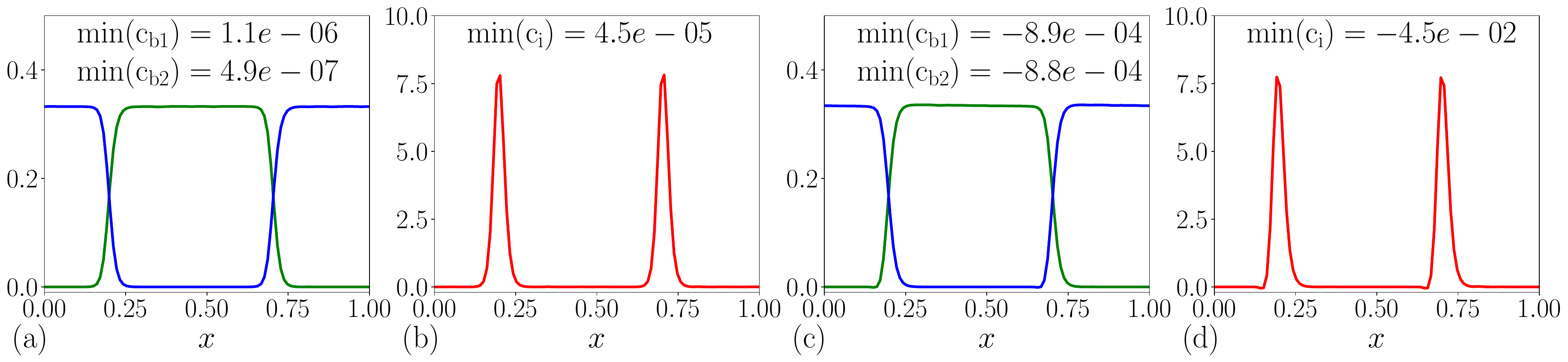}
    \caption{One-dimensional simulation of adsorption of surfactant onto a moving droplet interface, showing (a,b) final bulk and interfacial concentrations for $Pe_c=0.5$, (c,d) final bulk and interfacial concentration for $Pe_c=2$.}
    \label{fig:positive2}
\end{figure}

In this section, the robustness of the positivity criterion in Eqs. \eqref{eq:crossover} and \eqref{eq:positive} is evaluated. The setup used here is the same as in Section \ref{sec:adsorp-desorp}, but with a moving droplet. Here, the parameters are set as $r_{a,l}=1, r_{d,l}=0, D_{b,l}=1,$ and $D_i=1$, and the initial concentration values are set as $c_{b,l} = \phi_l$ and $c_i=0$ at time $t=0$ as shown in Figure \ref{fig:positive1} (a). After 1 flow-through time, the surfactant concentrations in the bulk and on the interface are shown in Figures \ref{fig:positive1} (b,c) for $Pe_c=1$, Figures \ref{fig:positive2} (a,b) for $Pe_c=0.5$, and Figures \ref{fig:positive2} (c,d) for $Pe_c=2$.
The minimum values of the concentration field are also reported in the plots. 
As excepted, the simulation with $Pe_c=2$ violates the positivity criterion, and therefore, negative values of the surfactant concentration are observed.


\subsection{Multi-dimensional simulations\label{sec:multiD}}

 In this section, the applicability of the proposed model in multi-dimensional cases is tested. Three simulations are presented in the subsequent sections: (a) surfactant adsorption onto a droplet - a two-dimensional version of the simulation presented in Section \ref{sec:adsorp-desorp}, and also acts as a verification case where the evolution of the bulk concentration is compared against the analytical solutions, (b) droplet oscillation with surfactants - a two-dimensional simulation of an oscillating droplet, where the effect of Marangoni stresses on the droplet oscillation is studied, and (c) droplet/bubble breakup in turbulence - a three-dimensional simulation of a forced two-phase turbulence from \citet{jain2025stationary}, where the robustness of the proposed model is verified in a complex turbulent flow in the presence of breakup and coalescence and the effect of the surfactants on the dispersed phase is studied.

\subsubsection{Surfactant adsorption}

\begin{figure}
    \centering
    \includegraphics[width=0.8\textwidth]{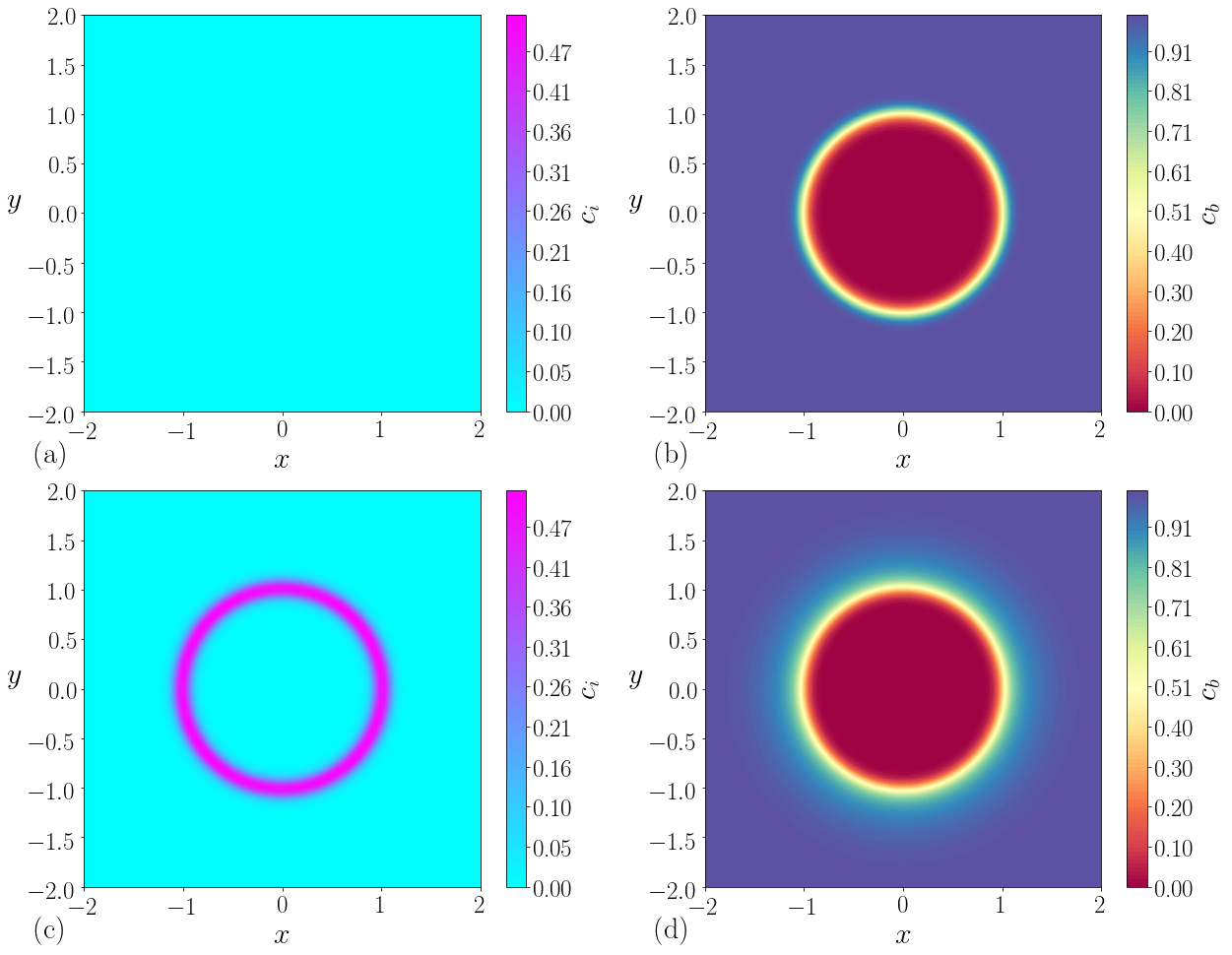}
    \caption{(a,b) Initial interfacial and bulk surfactant concentrations, respectively, and (c,d) interfacial and bulk surfactant concentrations at time $t=0.1$.}
    \label{fig:surf-adsorp}
\end{figure}

This canonical case was previously used by \citet{teigen2009diffuse} and \citet{muradoglu2008front} as a verification test. A initially clean stationary droplet of radius $1$ is placed at the center of a domain of size $4\times4$ [Figure \ref{fig:surf-adsorp}(a)], and the domain is discretized into a grid of size $100\times100$. The bulk droplet phase has a uniform initial surfactant concentration of $1$, initialized by setting $c_{b,1} = \phi$ (Figure \ref{fig:surf-adsorp}(b)). A simplified adsorption isotherm is adopted in this test case, for the sake of verification, as
\begin{equation*}
    \hat{j}_l = r_{a,l} \tilde{c}_{b,l},
\end{equation*}
and the parameters are set as $r_{a,1}=1, r_{a,2}=0, D_{b,l}=1$, and $D_i=1$. The surfactant adsorbs onto the interface, the interfacial concentration increases with time, and the bulk concentration reduces close to the interface due to the adsorption. The interfacial and bulk surfactant concentration are shown in Figure \ref{fig:surf-adsorp}(c,d) at time $t=0.1$. 

\begin{figure}
    \centering
    \includegraphics[width=0.45\textwidth]{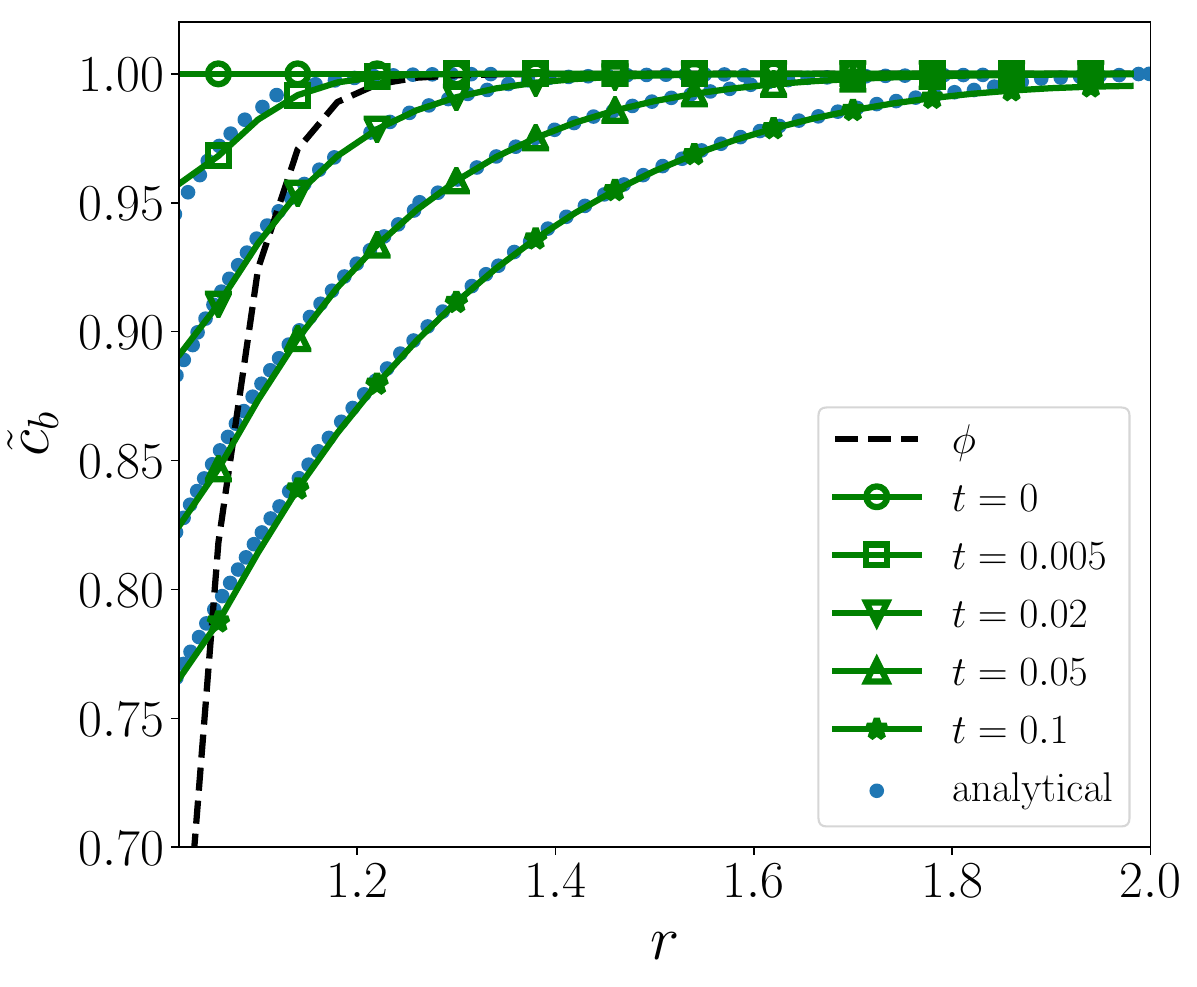}
    \caption{Local bulk concentration outside the drop at various times.}
    \label{fig:bulk-conc}
\end{figure}

The evolution of bulk surfactant concentration is governed by a heat equation with a sink at the interface location, given by
\begin{equation}
    \frac{\partial \tilde{c}_{b,l}}{\partial t} = \frac{D_{b,l}}{r} \frac{\partial }{\partial r} \left(r \frac{\partial \tilde{c}_{b,l}}{\partial r} \right),
\end{equation}

\begin{equation}
    \frac{\partial \tilde{c}_{b,l}}{\partial r} = - r_a \tilde{c}_{b,l},\hspace{1cm} \mathrm{at\ interface}
\end{equation}
and it was solved using a higher-order scheme to obtain a semi-analytical reference solution \citep{teigen2009diffuse} . We use this reference solution to compare against the accuracy of the proposed model in this work. Figure \ref{fig:bulk-conc} shows the local bulk concentration outside the drop at various times and compares it with the semi-analytical reference solution. The present method compares well with the reference solution, verifying the method.

\subsubsection{Droplet oscillation with surfactants}


In this section, we present a simulation of effect of surfactants on an oscillating droplet. The oscillating droplet without surfactants is a standard test case that is typically used to assess the accuracy of the method for capturing surface tension effects \citep{perigaud2005compressible,olsson2007conservative,ii2012interface,shukla2014nonlinear,garrick2017finite,jain2020conservative}. 
Consider a two-dimensional domain of dimensions $[-2,2]\times[-2,2]$, and the domain is discretized into a grid of size $100\times 100$. An initially clean ellipse-shaped droplet with semi-major and semi-minor axes of sizes $0.75$ and $0.4$, respectively, is placed in the center of the domain. The bulk droplet phase has a uniform initial surfactant concentration of $1$. Three simulations are performed, one (a) without surfactant (clean case), and two cases with surfactant with (b) $r_{a,1}=1, r_{a,2}=0,$ and $r_{d,l}=0$, and (c) $r_{a,1}=1, r_{d,1} = 0.5, r_{a,2}=0,$ and $r_{d,2}=0$. Other parameters are chosen to be $D_{b,l}=1, D_i=1, \rho_1=1000, \rho_2=1, \sigma_0=1, Ma=1$, and $\hat{c}_{i,\infty}=1$.

\begin{figure}
    \centering
    \includegraphics[width=\textwidth]{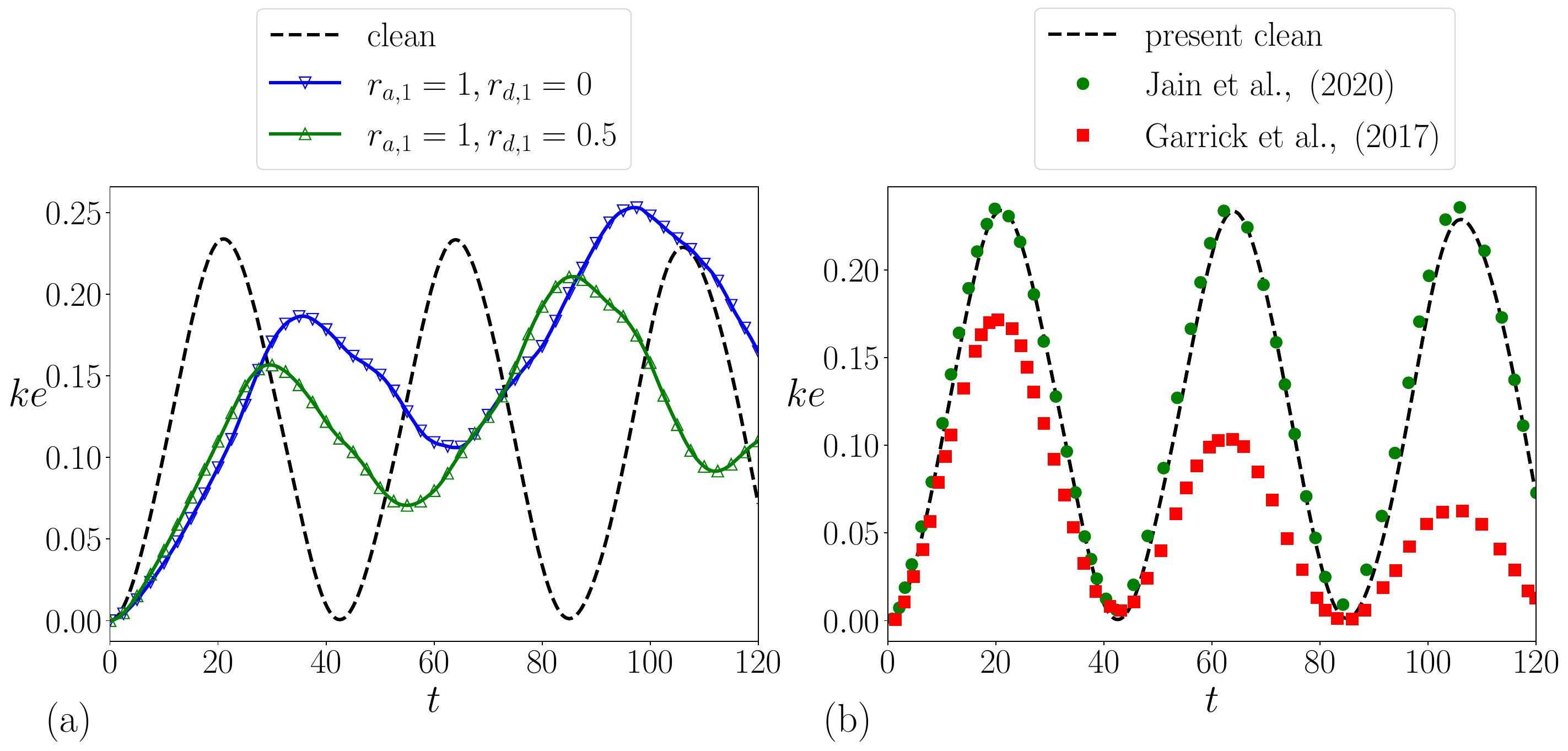}
    \caption{Kinetic energy of droplet oscillation (a) with and without surfactants, and (b) comparison with the previous studies \citep{jain2020conservative,garrick2017finite} for the clean case.}
    \label{fig:ke-surf}
\end{figure}

\begin{figure}
    \centering
    \includegraphics[width=0.8\textwidth]{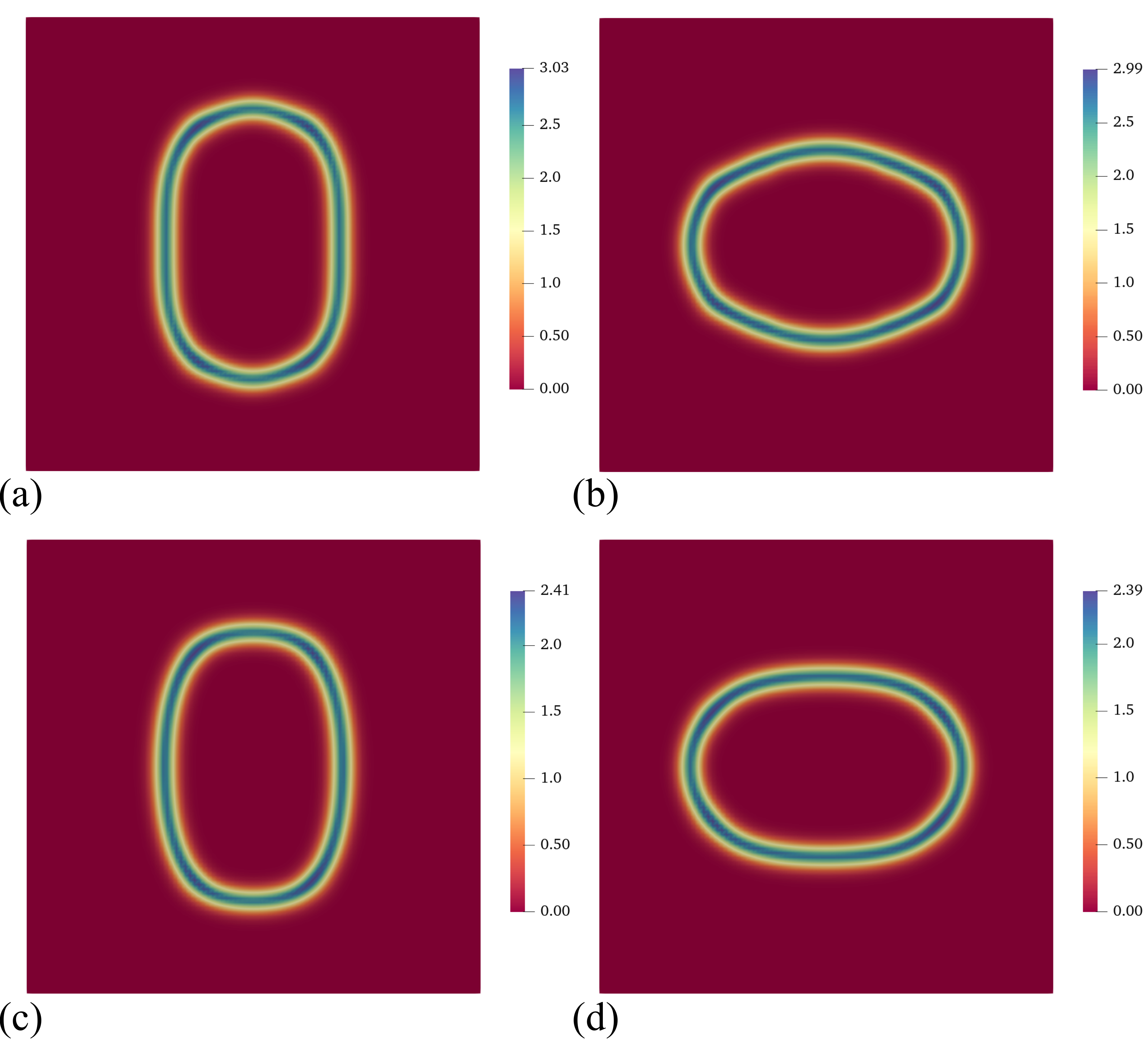}
    \caption{(a,b) Interfacial surfactant concentrations at time $t=60$ and $t=120$, respectively, for the case with $r_{a,1}=1, r_{a,2}=0,$ and $r_{d,l}=0$, and (c,d) interfacial surfactant concentrations at time $t=60$ and $t=120$, respectively, for the case with $r_{a,1}=1, r_{d,1} = 0.5, r_{a,2}=0,$ and $r_{d,2}=0$.}
    \label{fig:surf-reorg}
\end{figure}

Since the equilibrium shape of the droplet is a circle, surface tension forces deform the droplet toward its equilibrium shape, and this results in an oscillating droplet motion.
Figure \ref{fig:ke-surf} shows the kinetic energy in the domain as a function of time. Due to the exchange of energy between kinetic energy and surface energy, this plot is a signature of droplet oscillation. 
In the clean case, the droplet oscillates at its natural frequency throughout the simulation, as expected, and the kinetic energy is preserved throughout the simulation due to the use of non-dissipative schemes. The kinetic energy evolution for the clean case is shown in Figure \ref{fig:ke-surf} (b) and compared with the previous studies, which verifies the oscillation frequency and the non-dissipative nature of the present approach.
But with surfactants, the frequency of oscillation is modified.
The frequency of oscillation is reduced in both cases in the presence of surfactants compared to the clean case. This is because of the reduced value of surface tension coefficient in the presence of surfactants, and the frequency is related to the surface tension coefficient as
\begin{equation}
    f \propto \sqrt{\sigma}.
\end{equation}
Consequently, for case (b), the oscillation frequency is lower compared to case (a) because of the absence of desorption, which results in higher surfactant concentrations on the interface.  
In addition to the reduced oscillation frequency due to lower surface tension, the presence of Marangoni effects introduces non-periodic and higher modes of droplet oscillation as can be seen in Figure \ref{fig:surf-reorg}. 

\subsubsection{Droplet/bubble breakup in turbulence with soluble surfactants}


In this section, we present simulations of forced droplet/bubble-laden isotropic turbulence from \citet{jain2025stationary} in the presence of a soluble surfactant, and study the effect of surfactants on this stationary two-phase turbulence. This simulation illustrates the most challenging environment of transport of soluble surfactants in bubbles/drops undergoing breakup and coalescence in turbulence. 

The computational domain is of size $(2\pi)^3$ and is discretized into a grid of size $256^3$. This grid size was chosen such that it satisfies the scaling and resolution requirements for direct numerical simulations of two-phase turbulence, for both with and without surfactants, proposed in \citet{hatashita2025scalings}.
A single-phase simulation is first performed at a Taylor-microscale Reynolds number of $Re_{\lambda}\approx 87$ for about $t/\tau_e\approx 30$ with a controller-based linear forcing so that the stationary state is reached. Here $\tau=(2/3)k/\epsilon$ is the eddy turnover time, where $k=1$ is the turbulent kinetic energy and $\epsilon=0.4332$ is the dissipation. Then a spherical droplet of diameter $1$ is placed at the center of the domain along with the soluble surfactant inside the bulk of the droplet phase. The surfactant has a uniform initial concentration of $1$ inside the droplet when introduced. Once the droplet and the surfactant are introduced, the simulation is continued for another $40$ eddy-turnover time while being forced in the carrier phase to maintain a constant kinetic energy in the carrier phase. The details of the two-phase forcing can be found in \citet{jain2025stationary}. A similar simulation is performed without surfactants so that the effect of surfactants can be quantified. The properties chosen in the simulations are $D_i=0.2, D_{b,l}=0.2, r_{a,1}=1, r_{d,1} = 0, r_{a,2}=1,$ and $r_{d,2}=0, Ma=0.5, c_{\infty}=1, \rho_l=1$, and $\mu_l=2.033\times 10^{-3}$. The surface tension is chosen to be $\sigma=0.1620$, which results in an integral-scale Weber number \citep{jain2025stationary} of about $9.75$. 

\begin{figure}
    \centering
    \includegraphics[width=\linewidth]{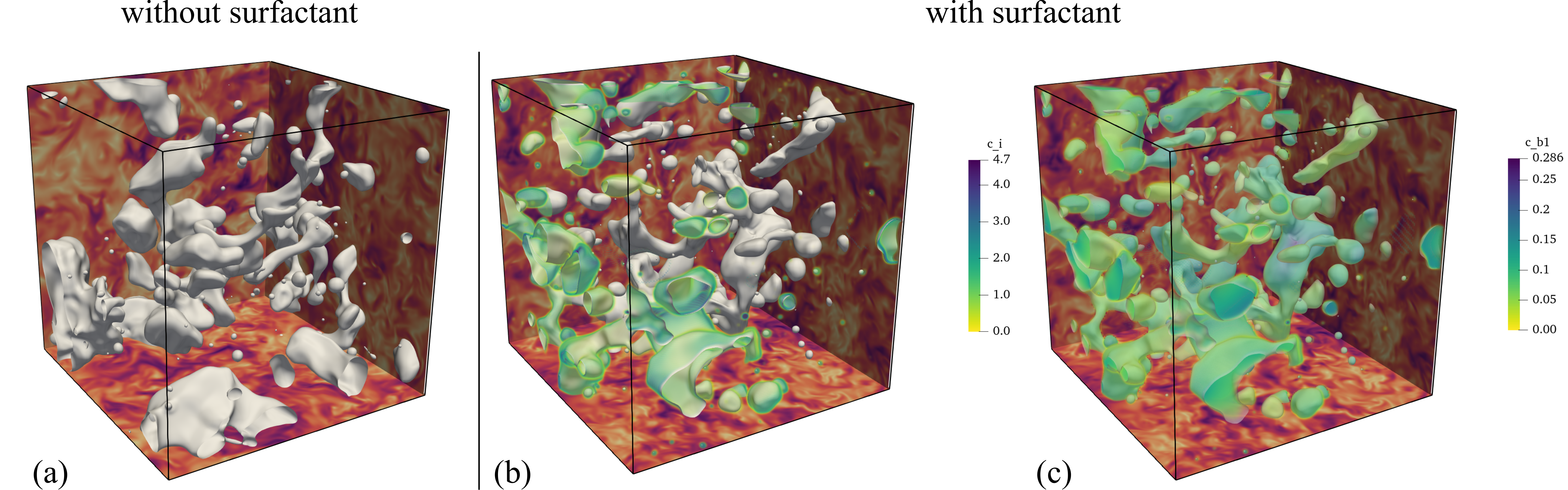}
    \caption{Bubble/droplet-laden isotropic turbulence (a) without surfactants, and (b,c) with soluble surfactants at $t/\tau_e\approx 4$, showing the dispersed phase and the velocity fields. For the cases with soluble surfactants, a volumetric rendering of interfacial surfactant concentration is shown in (b) and bulk surfactant concentration is shown (c).}
    \label{fig:bubbleHIT}
\end{figure}

Figure \ref{fig:bubbleHIT} shows the droplet/bubble-laden isotropic turbulence at $t/\tau_e\approx 4$ after the droplet and surfactant are introduced into the domain, with and without surfactants. The droplet undergoes breakup due to turbulence and then reaches a stationary state due to forcing. Throughout the simulation, interfacial surfactant is seen to be confined to the interface region and the bulk surfactant is confined to the droplet phase, proving the effectiveness of the model. In the presence of surfactants, more finer droplets can be seen in the domain when compared to the simulation without surfactants. This is expected because surfactants are known to prevent coalescence, resulting in finer droplets.

\begin{figure}
    \centering
    (a)
    \includegraphics[width=0.45\linewidth]{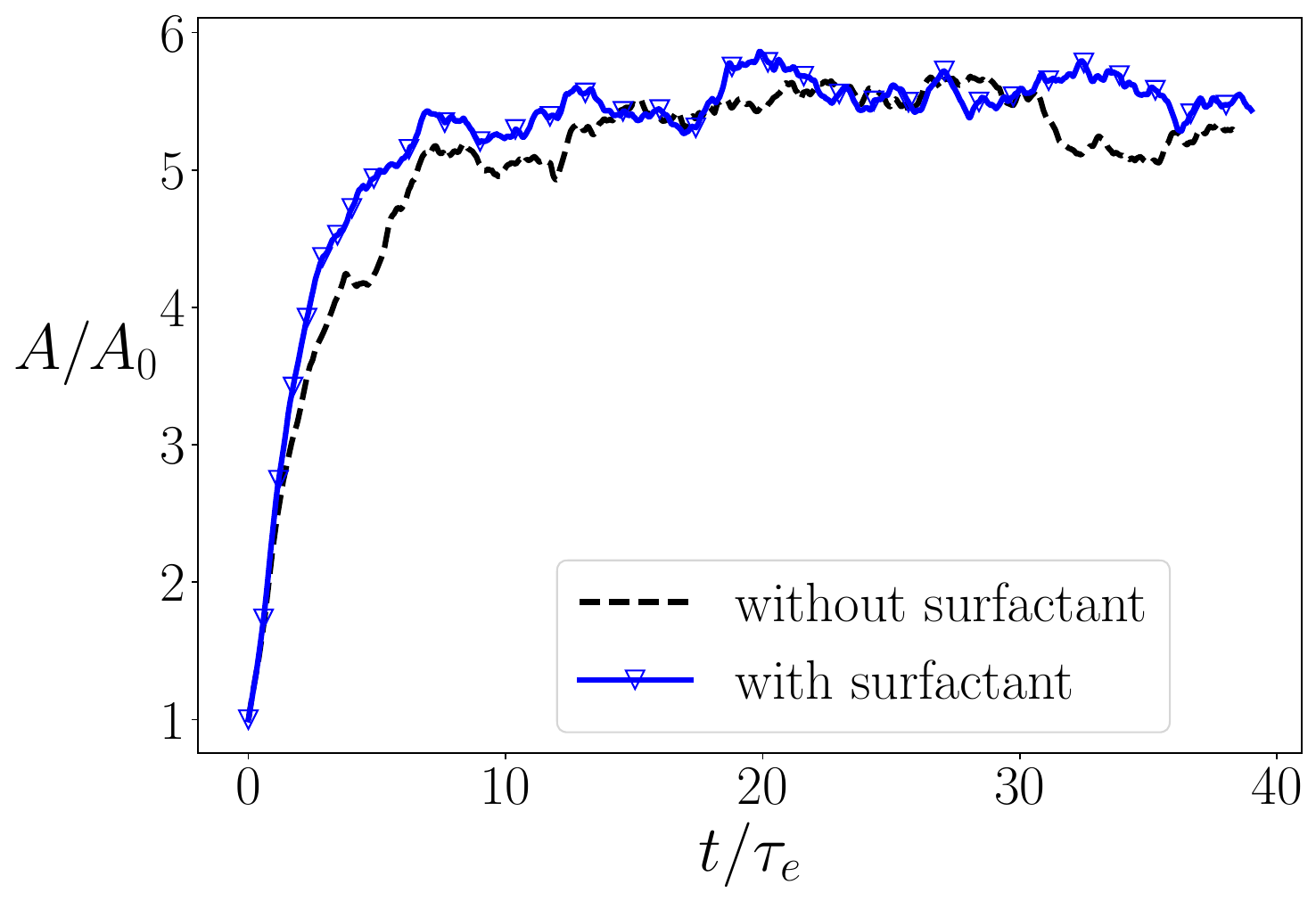}
    (b)
    \includegraphics[width=0.45\linewidth]{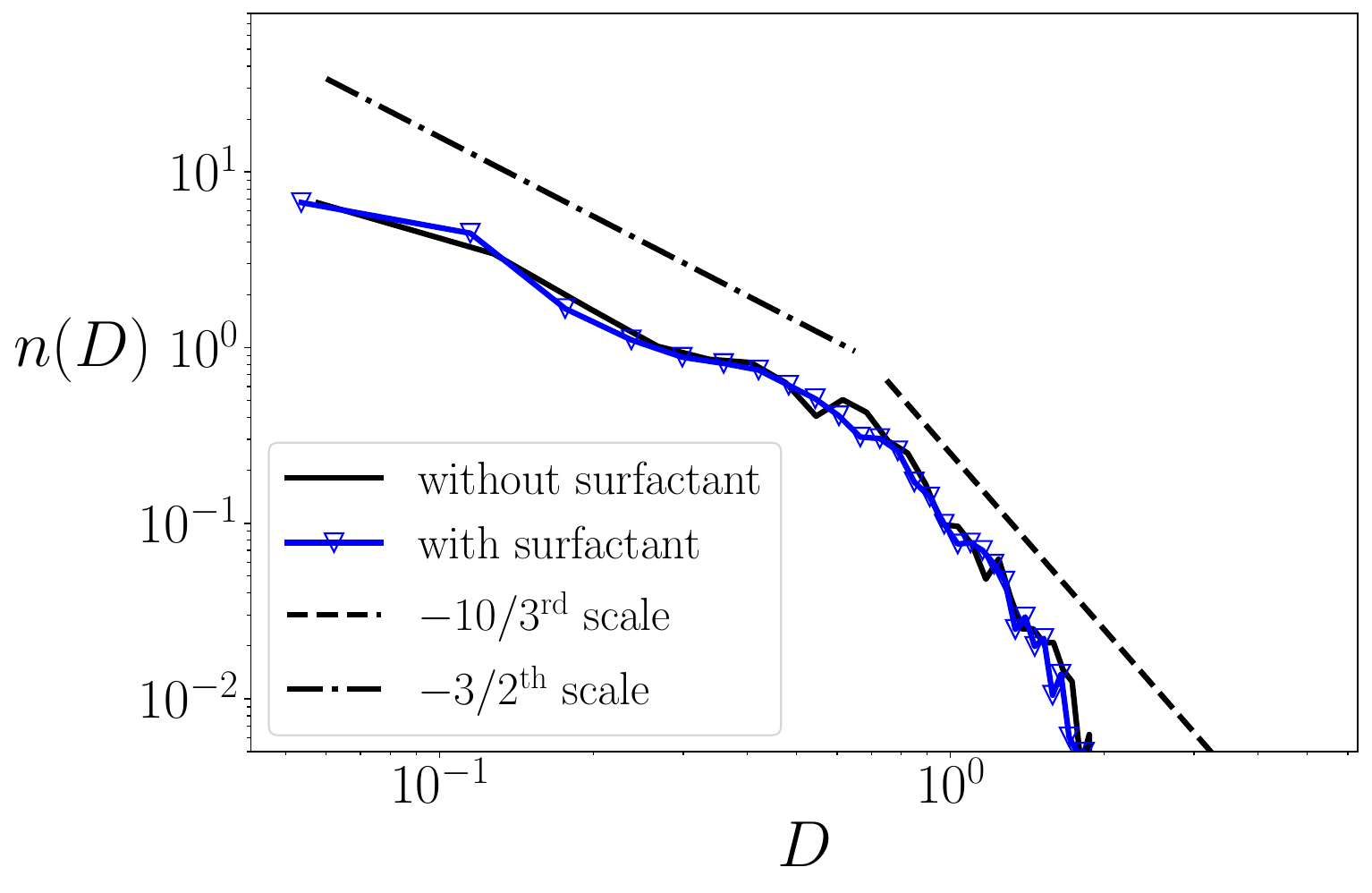}
    \caption{(a) Total interfacial area with and without surfactants. (b) Drop/bubble size distribution with and without surfactants.}
    \label{fig:area-size}
\end{figure}

To further quantify the effect of surfactants, total interfacial area in the domain is computed as a function of time and is shown in Figure \ref{fig:area-size} (a). Because of the coalescence inhibition in the presence of surfactants, the total interfacial area is higher compared to the case without surfactant. When time averaged (in stationary state), the total interfacial area $A/A_0$ is found to be $5.39$ without surfactants and $5.54$ with surfactants, which clearly shows the enhancement of interfacial area due to the addition of surfactants. 
Figure \ref{fig:area-size} (b) shows the drop size distributions computed using the volume-corrected flood-fill algorithm that accounts for the diffuse nature of the interface to recover accurate droplet volumes/sizes \citep{nathan2025accurate}. Slightly higher droplet count can be seen for smaller droplet sizes (around $D\approx 0.1$) in the presence of surfactants, due to the inhibition of coalescence, which results in enhancement of the interfacial area.

\section{Conclusion}


In this work, a model for the transport of soluble surfactants in two-phase flows is developed. The proposed model is solved with a second-order phase-field model with a non-dissipative central scheme; however, it also can be used with other interface-capturing methods. 
The method discretely conserves the total surfactant mass and results in positive surfactant concentrations, a physical-realizability (robustness) condition, provided the given positivity criterion is satisfied. The proposed method also does not allow artificial leakage of the surfactant between the interface and the bulk and between the two phases in the bulk.

The proposed model was used to simulate a wide variety of cases including adsorption/desorption of surfactant on/to a stationary and moving droplet. The model was verified to maintain the positivity of the surfactant concentration, and the accuracy of the model was verified by comparing it with analytical solutions for the adsorption process. 
Finally, the effect of soluble surfactants on an oscillating droplet and on complex droplet/bubble-laden isotropic turbulence was also studied.


\section*{Acknowledgments} 

This work was supported by the donors of ACS Petroleum Research Fund under Doctoral New Investigator Grant 69196-DNI9. S.~S.~J. served as Principal Investigator on ACS PRF 69196-DNI9.
S.~S.~J. also acknowledges partial financial support from the George W. Woodruff School of Mechanical Engineering at Georgia Institute of Technology and the generous computing resources from the DOE's ALCC awards (TUR147 \& BubbleLaden, PI: Jain). This research used supporting resources at the Argonne and the Oak Ridge Leadership Computing Facilities. The Argonne Leadership Computing Facility at Argonne National Laboratory is supported by the Office of Science of the U.S. DOE under Contract No. DE-AC02-06CH11357. The Oak Ridge Leadership Computing Facility at the Oak Ridge National Laboratory is supported by the Office of Science of the U.S. DOE under Contract No. DE-AC05-00OR22725. A preliminary version of this work has been published \citep{jain2023modeling} as the Center for Turbulence Research Annual Research Brief (CTR-ARB) and is available online\footnote{\href{https://web.stanford.edu/group/ctr/ResBriefs/2023/13_Jain.pdf}{https://web.stanford.edu/group/ctr/ResBriefs/2023/13\_Jain.pdf}}.

\section*{Appendix A: Alternate models}

Following the procedure used by \citet{jain2020conservative}, additional consistent terms can be derived for the proposed model (subscript $l$ is dropped here) as
\begin{flushleft}
\textbf{Model A}:\\
\end{flushleft}
\begin{equation}
\frac{\partial c_s}{\partial t} + \vec{\nabla}\cdot(\vec{u} c_s ) = \vec{\nabla} \cdot \left[D \left\{\vec{\nabla}c_s - \frac{2(0.5 - \phi) \vec{n} c_s}{\epsilon} \right\}\right] + \vec{n} \cdot \vec{\nabla} \left(\vec{\nabla}\cdot \vec{a}\right) \hat{c}_s + \hat{j} \delta_s ,
\label{equ:interface_surfactant_consistent_model1}
\end{equation}
\begin{equation}
\frac{\partial c_b}{\partial t} + \vec{\nabla}\cdot(\vec{u} c_b) = \vec{\nabla} \cdot \left[D \left\{\vec{\nabla}c_b - \frac{(1 - \phi) \vec{n} c_b}{\epsilon} \right\}\right] + \vec{\nabla}\cdot \left(\tilde{c}_b\vec{a}\right) - \hat{j} \delta_s.
\label{equ:bulk_surfactant_consistent_model1}
\end{equation}
However, note that the additional consistent terms are in a non-conservative form; therefore, care must be taken in using this model form. Using a model form based on \citet{teigen2009diffuse}, we can also write the proposed model as
\begin{flushleft}
\textbf{Model B}:\\
\end{flushleft}
\begin{equation}
\frac{\partial c_s}{\partial t} + \vec{\nabla}\cdot(\vec{u} c_s ) = \vec{\nabla} \cdot \left(D \delta_s \vec{\nabla} \hat{c}_s \right) + \vec{n} \cdot \vec{\nabla} \left(\vec{\nabla}\cdot \vec{a}\right) \hat{c}_s + \hat{j} \delta_s ,
\label{equ:interface_surfactant_consistent_model2}
\end{equation}
\begin{equation}
\frac{\partial c_b}{\partial t} + \vec{\nabla}\cdot(\vec{u} c_b) =  \vec{\nabla} \cdot \left(D \phi \vec{\nabla} \tilde{c}_b \right)
+ \vec{\nabla}\cdot \left(\tilde{c}_b\vec{a}\right) - \hat{j} \delta_s.
\label{equ:bulk_surfactant_consistent_model2}
\end{equation}
The additional terms in Eqs. \eqref{equ:bulk_surfactant_consistent_model1} and \eqref{equ:bulk_surfactant_consistent_model2} are the same as the ones introduced by \citet{jain2020conservative}. The advantage of \textbf{Model B} is that it does not assume a perfect equilibrium interface; however, it requires computation of $\hat{c}_s$ in the diffusion term, which requires division by $\delta_s$ and could be non-robust.

\bibliographystyle{model1-num-names}
\bibliography{two_phase}

\end{document}